\pdfoutput=1

\documentclass[11pt]{article} 

\usepackage[]{EACL2023}
\usepackage{microtype}
\usepackage{times}
\usepackage{paralist}
\usepackage{graphicx}
\usepackage{wrapfig}
\usepackage[inline]{enumitem}
\usepackage{amssymb,multirow}
\usepackage{natbib}
\usepackage{booktabs}
\usepackage{comment}
\usepackage{subcaption}

\title{Fréchet Distance for Offline Evaluation of  \\ Information Retrieval Systems with Sparse Labels}

\author{Negar Arabzadeh \\
University of Waterloo \\
  \texttt{narabzad@uwaterloo.ca} \\\And
Chalres L. A. Clarke  \\
University of Waterloo \\
\texttt{claclark@uwaterloo.ca}}

\begin{document}
\maketitle
\begin{abstract}
{

The rapid advancement of natural language processing, information retrieval (IR), computer vision, and other technologies has presented significant challenges in evaluating the performance of these systems. One of the main challenges is the scarcity of human-labeled data, which hinders the fair and accurate assessment of these systems. In this work, we specifically focus on evaluating IR systems with sparse labels, borrowing from recent research on evaluating computer vision tasks.
taking inspiration from the success of using Fréchet Inception Distance (FID) in assessing text-to-image generation systems. We propose leveraging the Fréchet Distance to measure the distance between the distributions of relevant judged items and retrieved results. Our experimental results on MS MARCO V1 dataset and TREC Deep Learning Tracks query sets demonstrate the effectiveness of the Fréchet Distance as a metric for evaluating IR systems, 
particularly in settings where a few labels are available.
This approach contributes to the advancement of evaluation methodologies in real-world scenarios such as the assessment of generative IR systems.
}

\end{abstract}

\section{Introduction}

With the rapid advancement of technologies in fields such as natural language processing, natural language generation, computer vision, and information retrieval (IR), evaluating the performance of these systems is becoming increasingly challenging~\cite{gatt2018survey,hashimoto2019unifying,celikyilmaz2020evaluation,yang2020evaluation}.
We must develop new metrics, benchmarks, and evaluation protocols that are specifically tailored to the unique characteristics of the systems considering the rapid changes in system architecture, training data, and model configurations~\cite{theis2015note}. 
In many cases, obtaining high-quality labeled data that accurately represents the complexity of real-world scenarios can be expensive, time-consuming, or even impractical. This scarcity of labeled data adds to the limitations of conducting extensive evaluations and may lead to biased or incomplete assessments~\cite{arabzadeh2022shallow}.

Offline evaluation poses a significant challenge due to the sparsity of labeled data~\cite{clarke2023preference,clarke2020offline,xie2020preference,arabzadeh2023quantifying,arabzadeh2023adele}. This challenge is particularly prominent in datasets like MS MARCO, a widely used benchmark for ad hoc retrieval reserach~\cite{nguyen2016ms,arabzadeh2021ms,mackenzie2021sensitivity,arabzadeh2024adapting,Huo_2023} in which, the majority of queries are annotated with only one relevant judged document. However, to suit the dataset for effective traininig of deep learning models, a high number of queries are judged, resulting in sparse labels. Consequently, most queries have only one relevant judgment, while the relevance of the remaining documents remains unknown. Other researchers have shown that there are potentially relevant documents that are as good as, or even better than, the judged queries \cite{qu2020rocketqa,arabzadeh2022shallow}. Given the sparsity of ground truth labels, it is crucial to recognize the challenges involved in distinguishing between rankers when the differences in performance are small \cite{yan2022human}. The limited labeled data for retrieved documents introduces noise, making it challenging to definitively determine which ranker is performing better \cite{cai2022hard}.
The incomplete judgments can introduce problems in evaluations, as they do not capture the full range of relevant documents \cite{aslam2006statistical,carterette2007hypothesis}. This issue becomes even more pronounced in generative-based tasks. It is impractical to reassess the generated results, such as images or text, with each system run due to their non-deterministic nature \cite{theis2015note,harshvardhan2020comprehensive}. 

Evaluating a generative system's performance based on the similarity of generated content to sparsely labeled data remains one of the most effective approaches in many generative-based NLP and computer vision benchmarks and tasks~\cite{soloveitchik2021conditional,heusel2017gans,obukhov2020quality,dimitrakopoulos2020wind,bertscore}. Particularly in the evaluation of text-to-image generation task, the Fréchet Inception Distance (FID), has gained recognition for showing high robustness and correlation with human judgements~\cite{heusel2017gans,saharia2022photorealistic,yu2022scaling}. FID compares the distribution of generated images across a set of prompts to the distribution of target images across the same set of prompts. 
To compute FID, features of ground truth images and generated images are extracted from both sets, and multivariate Gaussian distributions are fitted to these features. The Fréchet Distance ($\textit{FD}$), which quantifies the similarity between two probability distributions, is then computed based on the fitted Gaussian distributions. A lower FID score indicates a higher similarity between the distributions, indicating that the generated images closely match the real images in terms of their visual features.

In this paper, we shed light on how evaluating generated results is similar to assessing the quality of retrieved results with sparse labels in an ad hoc retrieval setting. Most benchmarks for both tasks have quite sparse labels i.e., not all the items are judged and  while there are a few annotations available for some of the candidates, there can be other unjudged relevant items available. While labelling more data is expensive for both tasks, there could be more than one correct answer in both tasks. In this work, we mimic an Information Retrieval system with sparse relevance judgements as a generation task where the ground truth targets are sparse. Due to the success of  FID in evaluating the quality of generated images, especially for generative adversarial networks \cite{gafni2022make,saharia2022photorealistic,yu2022scaling,khan2020realistic,alonso2019adversarial}, we explore if we can quantify the quality of retrieved documents in an ad hoc retrieval system through \textit{Fréchet Distance}.
In the context of IR evaluation, we can analogously consider the relevant judged items as the ground truth set and the retrieved items as the set of generated items. Our objective is to extract features from both sets, the relevant judged items and the retrieved results, and investigate whether metrics such as the Fréchet Distance can effectively capture the quality of the retrieved results with respect to the ground truth labels in IR systems. 

We study the following Research Questions:
\begin{itemize}[wide, labelwidth=!, labelindent=0pt]
\itemsep0em 
    \item RQ1. Can the Fréchet Distance effectively evaluate IR systems with sparse labels? 
    \item RQ2. Can the Fréchet Distance effectively evaluate IR systems with comprehensive labels?
    \item RQ3. Can the Fréchet Distance effectively evaluate the quality of IR systems when the retrieved results are not labelled?
    \item RQ4. How well correlated are the performance of IR systems, as measured by the Fréchet Distance vs.\ and traditional IR metrics?
    \item RQ5. How robust is the Fréchet Distance for evaluating IR systems with respect to the feature extraction methods used to represent both the ground truth and retrieved items?
\end{itemize}

We conduct our experiments by assessing different retrieval pipelines on the MS MARCO V1 Dev dataset, which has extremely sparse labels, as well as the TREC Deep Learning Track 2019 and 2020 datasets, which have more complete labels \cite{nguyen2016ms,craswell2020overview,dl20}. Our study demonstrates the effectiveness of the Fréchet Distance as a metric for quantifying the performance of IR systems especially when the ground truth labels are sparse.


\section{Fréchet Distance for IR evalaution}

\subsection{Fréchet Distance}
The Fréchet distance is a measure of dissimilarity between two curves or trajectories and has shown to be useful in numerous applications including computational geometry, computer graphics, bioinformatics and robotics \cite{alt2009computational,alt1995computing,alt2001matching,jiang2008protein,gheibi2014minimum}. To understand the Fréchet distance, let us consider two curves (or trajectories or paths): $A$ and $B$. The Fréchet distance between $A$ and $B$ could be exemplified as  measuring the minimum leash length required by a dog walking along a path $A$ while its owner walks along path $B$, with both the dog and owner potentially traversing their respective paths at different speeds~\cite{DBLP:journals/corr/abs-cs-0703011,eiter1994computing}. The leash cannot be shortened or lengthened during the walk. The definition is symmetric i.e., the Fréchet distance would be the same if the dog were walking its owner.
Given two curves, $A$ and $B$, represented as sequences of points in a metric space, the Fréchet distance, denoted as $F(A, B)$ is computed as:
\begin{equation}
\small    F(A, B) = inf_{\alpha,\beta} max_{t\in[0,1]} d(A(\alpha(t)), B(\beta(t)))
\end{equation}
where $A$ and $B$ are continues maps from $[0,1] $ to  metric space and $\alpha$ and $\beta$ are reparameterizations of the unit interval $[0,1]$ i.e. they are continuous, non-decreasing, surjection functions. The requirement of non-decreasing reparameterizations, $\alpha$ and $\beta$, ensures that neither the dog nor its owner can backtrack along their respective curves.
The parameter $t$ as represents the progression of time, consecutively $A(\alpha(t))$ and and $B(\beta(t))$ represent the position of the dog  and the dog's owner at time $t$ (or vice versa). The distance $d$ between $A(\alpha(t))$ and $B(\beta(t))$ corresponds to the length of the leash between them at time $t$. By considering the \textit{infimum} over all potential reparameterizations of the unit interval $[0,1]$, we select the specific paths where the maximum leash length is minimized. 

Apart from quantifying the dissimilarity between curves, the Fréchet distance can also serve as a measure to assess the disparity between probability distributions~\cite{heusel2017gans}.Given we have two normal univariate distributions, $X$ and $Y$, Fréchet Distance $(\textit{FD})$  is given as:
\begin{equation}
   \small \textit{FD}(X,Y)=(\mu_{X} - \mu_{Y})^2 + (\sigma_{X} - \sigma_{Y})^2
\end{equation}
Where $\mu$ and $\sigma$ are the mean and standard deviation of the normal distributions, respectively.

\subsection{Fréchet Inception Distance}
In computer vision, the Inception V3 model pre-trained on the Imagenet dataset is employed to generate feature vectors to be approximated by multivariate normal distribution \cite{DBLP:journals/corr/SzegedyVISW15}. As such, the Fréchet Inception Distance (FID) for a multivariate normal distribution is computed as:
\begin{equation}
\label{fid}
  \footnotesize  \textit{FID}(X,Y)= || \mu_{X} - \mu_{Y} || ^2 - Tr( \Sigma_X + \Sigma_Y - 2\sqrt{\Sigma_X \Sigma_Y})
\end{equation}
In this equation, $X$ and $Y$ represent two distributions derived from two sets of embeddings. These embeddings correspond to real images and generated images, respectively, and are obtained from the Inception model. 
The vectors $X$ and $Y$ have magnitudes $\mu_X$ and $\mu_Y$, respectively. The trace of the matrix is denoted as $Tr$, while $\Sigma_X$ and $\Sigma_Y$ represent the covariance matrices of the vectors.

\subsection{Fréchet Distance for IR}
Let us assume $C$ is a collection of items and $Q = \{q_1, q_2, …, q_n\}$ is a set of $n$ queries, where each query $q_i$ has a set of relevant judged items $R_{q_i}$. We define $R_Q$ as a set of relevance judged items for queries in $Q$, where $R_Q = \{ d | d \in R_{q_i} , q_i \in Q\}$. Furthermore, we can obtain the top-$k$ retrieved items by a retrieval system $M$ from $C$ for a given query $q$ as $M_k(q,C) = D^k_q$, where $D^k_q$ is a set of the top-$k$ most relevant retrieved items for query $q$, i.e., $D^k_q = \{d_1^q, d_2^q, …, d_1^k\}$.
Given $\mathbb{V}$ as a function that maps any retrieved item to a $ p$-dimensional embedding space, where $p$ is usually in the order of  a few hundred, 
 we can embed all the retrieved items and relevant judged items through $\mathbb{V}$. For instance, $\mathbb{V}(d_1)$ returns a $p$-dimensional vector embedding for document $d_1$.
To apply Fréchet Distance for assessing the quality of the IR system $M$, we measure $\textit{FD}_Q^M$ as follows on query set $Q$:
\begin{equation}
\small \textit{FD}_Q^{M_k} = \textit{FD}\Bigl(\{\mathbb{V}(R_Q)\}, \{\mathbb{V}(M_k(Q,C))\}\Bigr)
\end{equation}
Here, $\textit{FD}$ is the Fréchet Distance (Eq. \ref{fid}) measures the distance between the distribution of the set embeddings of the relevant judged items $\{\mathbb{V}(R_Q)\}$ and those of the retrieved items $ \{\mathbb{V}(M_k(Q,C))\}$. The lower $\textit{FD}_Q^{M_k}$ represents the retrieved items to have higher similarity with the relevant judged items and thus the better performance of the retrieval system $M$ on the query set $Q$.

\section{Experimental Setup}
\label{expset}

In this section, we describe the general settings of our experiments including datasets, the traditional IR metrics, retrieval methods and the pre-trained language models we used to embed the documents.

\subsection{Dataset and Query sets} We perform experiments on the MS MARCO passage retrieval collection V1
\footnote{\url{https://microsoft.github.io/msmarco/}}
, which includes over 8.8 million passages~\cite{nguyen2016ms}. First, in section \ref{results1}, we experiment on the 6980 queries in MS MARCO small dev set, which are sparsely labelled. The majority of the queries in this set (over 94\%) have only one relevant judged document per query. Second, in Section \ref{results2}, we experiment on the TREC Deep Learning (DL) track 2019
\footnote{\url{https://microsoft.github.io/msmarco/TREC-Deep-Learning-2019.html}}
and 2020
\footnote{\url{https://microsoft.github.io/msmarco/TREC-Deep-Learning-2020.html}}
to study how varying and extending the relevance judgments would affect the evaluation process \cite{dl20,craswell2020overview}. The difference between the two query sets is that while the MS MARCO dev set has a higher number of queries (6980) judged, with mostly one relevant document per query, it leaves us with no extra information about the unannotated documents.  On the other hand, the TREC DL tracks have fewer queries judged (97), but each query has a comprehensive set of judgments with multi-level judgments ranging from 0-4, indicating the degree of relevance.

We compare the results of the $\textit{FD}$ score with  the official traditional IR evaluation metrics of each benchmark, i.e., MRR@10 for MS MARCO and nDCG@10 for TREC Deep Learning tracks.

\subsection{Retrieval models} 
\label{ret}
To conduct experiments on MS MARCO dev set, we consider a set of 12 retrieval methods that are well-distinguished for their efficiency or effectiveness, ranging from traditional high-dimensional bag-of-word sparse retrievers to more recent dense retrievers well as trained high-dimensional sparse models, which are representative of novel retrieval methods developed over the past five years.
Specifically, we consider BM25 as the representative of the sparse retrievers standalone as well as applying BM25 to expanded documents through DeepCT and DocT5Query document expansion methods \cite{bm25,nogueira2019doc2query,nogueira2019document,dai2019context}. We include a set of dense retrievers including RepBERT \cite{zhan2020repbert}, ANCE \cite{ANCE}, Sentence-BERT (SBERT) \cite{sbert}, COLBERT \cite{khattab2020colbert} and COLBERT-V2 \cite{colbertv2}. We also employ the more recently proposed high dimensional learnt sparse retrievers, UniCOIL and SPLADE \cite{Splade,lin2021few}. Furthermore, we consider hybrid retrievers \cite{lin2021batch} that fuse the retrieved items from BM25 and dense retrievers, to cover a variety of retrievers and assess the ability of $\textit{FD}$ to quantify the quality of retrieval fairly. We note that we employ some of the retrieval models from Pyserini\footnote{\url{https://github.com/castorini/pyserini}} 
\cite{Lin_etal_SIGIR2021_Pyserini} and some of the others from the paper's original GitHub repository.  
For more information about each of the retrieval models, we kindly refer to the original papers of each method.

For our experiments with the TREC DL19 and DL20 query sets, we took the submitted runs for each track from the NIST website\footnote{\url{https://trec.nist.gov/}}. Our experiments compare the results when assessing with Fréchet distance as well as nDCG@10 for 37 submitted runs to TREC DL2019 and 59 submitted runs to TREC DL 2020. These runs cover a comprehensive set of retrieval pipelines, typically with from sparse and/or dense retrieval as a retrieval first stage followed by one or more neural re-ranking stages~\cite{craswell2020overview,dl20}. 

\subsection{Embeddings}
To examine the robustness of $\textit{FD}$ on IR systems, we perform experiments using two different types of transformer-based contextualized models to embed the documents and extract their features.  We employ a general-purpose DistilBERT \cite{sanh2019distilbert} to obtain the documents embeddings\footnote{\url{https://bit.ly/3Oq39IB}} as well as fine-tuned pre-trained language models on MS MARCO\footnote{\url{https://bit.ly/3On7D2B}} \cite{sbert}. Both models were adapted from hugging face. We note that unless we explicitly mention (Section \ref{robustnessembd}) all the results are reported with the first model, i.e., the DistilBERT model that was fine-tuned on MS MARCO. 
We believe that by exploring different document representations, we may better understand the influence of document quality on the utilization of $\textit{FD}$ for evaluating IR systems. 

\section{Assessment with Sparse labels}
\label{results1}
We are interested in investigating how $\textit{FD}$ can assess the performance of different retrievers when there are only sparse labels available i.e., on 6980 queries from MS MARCO small dev set.  We present the performance of the 12 retrieval methods, including the sparse to dense retrievers, sparse retrievers with learned representations, and hybrid retrievers that were 
introduced in Section \ref{ret} in terms of MRR@10 as well as measuring the Fréchet Distance between two sets of retrieved items and relevant judged items on the cut-offs of 1 and 10 in Table \ref{mainresults}.  

\begin{table}[] 
\label{mainresults}
\caption{Performance of different retrievers in terms of MRR@10 as well as Fréchet distance $\textit{FD}$ on MS MARCO dev set. A smallest Fréchet distance corresponds to better performance. }
\scalebox{0.78}{
\begin{tabular}{llrrr}
\hline\hline
Category & Method & \multicolumn{1}{l}{MRR@10} & \multicolumn{1}{l}{$\textit{FD}@1$} & \multicolumn{1}{l}{$\textit{FD}@10$} \\ \hline
\multirow{3}{*}{Sparse} & BM25 & 0.187 & 7.446 & 4.410 \\ 
 & DeepCT & 0.242 & 1.453 & 2.354 \\ 
 & DocT5 & 0.276 & 3.047 & 2.050 \\ \hline
\multirow{5}{*}{Dense} & RepBERT & 0.297 & 1.881 & 1.223 \\ 
 & ANCE & 0.330 & 1.529 & 0.995 \\ 
 & SBERT & 0.333 & 1.387 & 1.008 \\ 
 & ColBERT & 0.335 & 1.456 & 0.980 \\ 
 & ColBERT V2 & 0.344 & 1.453 & 0.982 \\ \hline
\multirow{2}{*}{\begin{tabular}[c]{@{}l@{}}Trained\\  Sparse\end{tabular}} & UniCOIL & 0.351 & 1.387 & 0.980 \\ 
 & SPLADE & 0.368 & 1.328 & 0.964 \\ \hline
\multirow{2}{*}{\begin{tabular}[c]{@{}l@{}}Hybrid\\ (BM25)\end{tabular}} & ColBERT-H & 0.353 & 1.494 & 0.973 \\ 
 & ColBERT V2 -H & 0.368 & 1.464 & 0.998 \\ \hline\hline
\end{tabular}}

\label{mainresults}
\end{table}

The results for $\textit{FD}@1$ and $\textit{FD}@10$ demonstrate the ability of $\textit{FD}$ to quantify the performance of retrievers. For example, for the BM25 retriever, $\textit{FD}@1$ is measured as 7.446 and $\textit{FD}@10$ as 4.410. However, for a neural retriever like ColBERT, which has shown superior performance to BM25 on various benchmarks \cite{colbertv2,khattab2020colbert,DBLP:journals/corr/abs-2104-08663}, the $\textit{FD}$ values are reported as 1.456 and 0.980 for $\textit{FD}@1$ and $\textit{FD}@10$, respectively. This indicates that $\textit{FD}$ can effectively pickout the \textit{better} retriever, particularly when there is a significant difference between their performances. On the other hand, when the performance of two retrievers is quite similar, such as in the case of ColBERT vs. ColBERT V2, it becomes more challenging for evaluation metrics to assess their performance 
. For instance, while MRR@10 for ColBERT vs. ColBERT V2 is reported as 0.334 vs. 0.343, $\textit{FD}@10$ for the two retrievers is reported as 0.980 and 0.982. Therefore, as expected, the discriminative power of $\textit{FD}$ decreases when it becomes harder to distinguish between retrievers.
However, It is important to acknowledge that due to the sparsity of ground truth labels, previous research has indicated that distinguishing between rankers becomes challenging when the differences are small. In such cases,
the noise introduced by limited labeled data for retrieved documents makes it difficult to definitively determine which ranker is performing better \cite{qu2020rocketqa}. In fact \citet{arabzadeh2022shallow} showed that such a small difference in MRR@10 is not a strong indicator of which retrieval method is able to address the queries better since they might have surfaced other \textit{unjudged relevant items}. They showed that ordering of the rankers solely based on MRR and incomplete relevance judgement is not reliable.  Based on the results in Table \ref{mainresults} and their comparison with MRR@10, we can conclude that in response to \textbf{RQ1}, we observe that Fréchet Distance can effectively evaluate IR systems.

\begin{figure}[!t]
  \centering 
  \includegraphics[clip, trim=12.8cm 2.3cm 9.8cm 2.4cm,scale=0.7]{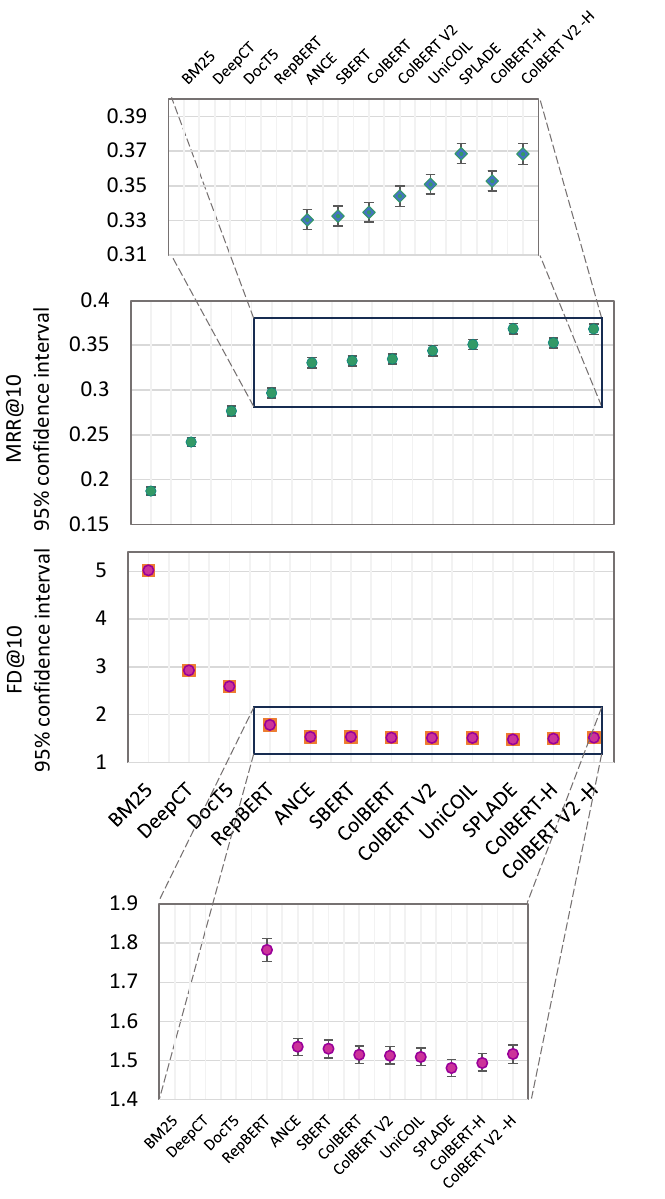}
  \caption{%
  	Performance of bootstrap sampling (N=1000) of queries in MS MARCO dev set in terms of MRR@10 and $\textit{FD}@10$ for the 12 different retrieval methods. 
  }
  \label{fig:distribution}
\end{figure}

\begin{figure*}[]
  \centering 
  \includegraphics[clip, trim=4cm 2cm 4.5cm 3.5cm,scale=0.95]{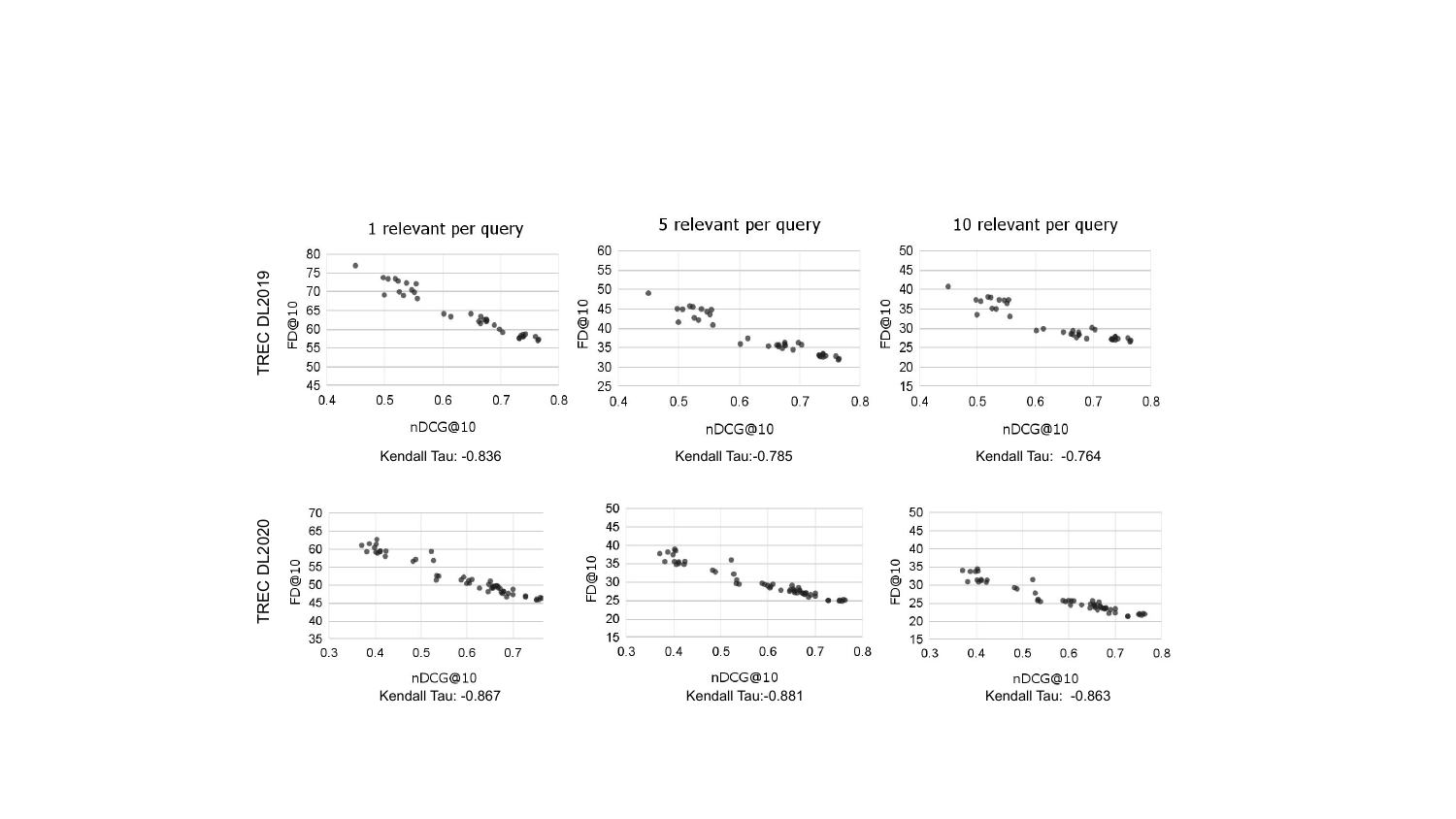}
  \caption{%
  	Performance of all the submitted runs to TREC DL 2019 (first row) and TREC DL 2020 (second row). In each sub-figure, X-axis and Y-axis indicate nDCG@10 and $\textit{FD}@10$ respectively. $\textit{FD}@10$ was measured with 1,5 and 10 relevant items per query in the sub-figures in the first, second and third columns respectively.    }
  \label{fig:trec}
\end{figure*}

To examine the robustness of the $\textit{FD}$ in the context of IR assessment, and to evaluate the generalizability of the method across different subsets of queries, we employ a bootstrap sampling \cite{johnson2001introduction,efron2003second} from the MSMARCO dev set for $N=1000$ times. This would allows us to investigate whether the results obtained in the previous section were influenced by the data or if they can be reliable.
The results are visualized in Figure \ref{fig:distribution}, in which we present the mean and empirical 0.95\% confidence interval for each retriever across the 1000 query sets in terms of MRR@10 and $\textit{FD}@10$. It is important to note that for the MRR plot, a higher position on the plot indicates better performance, while for the $\textit{FD}$ plot, a lower position indicates better performance.
The findings confirm that despite considering different sample sets, we observe a consistent pattern and similarity in the performance trends. 

\section{Assessing with Comprehensive labels}
\label{results2}

In this section, we investigate the performance of the Fréchet Distance in evaluating IR systems when the labels are not sparse and we have more complete labels.  We conduct experiments using the runs submitted to TREC DL 2019 (37 runs) and TREC DL 2020 (59 runs). Unlike the MS MARCO dev set which on average each query has 1.06 judged documents, the queries in TREC DL tracks on average have over 210 judged documents per query assessed with four different levels of relevance including ``not relevant'', ``related'', ``highly relevant'', and ``perfectly relevant'' 
\citet{craswell2020overview}.
We notice that the number of judged relevant items per query in these benchmarks varies a lot. Due to the TREC-style judgment criteria, only the top few retrieved items from all submitted runs were judged. Depending on the overlap between the top retrieved items from different runs, the number of relevant judged items per query may vary. When applying $\textit{FD}$ with an imbalanced number of relevant judged items per query, it can introduce biases in the ground truth distribution and potentially lead to problems in evaluation. To address this issue, we balanced the number of relevant judged items per query by limiting them to a maximum of 1, 5, and 10 relevant judged items per query i.e., we randomly select $K$ relevant items from the pool of relevant judged documents for the query of interest. We first randomly select from the most relevant level i.e., level 3 which are perfectly relevant documents and then when there is not a sufficient number of perfectly relevant documents, we move on to highly relevant level and randomly choose from that grade. This experiment also allows us to examine how the sparsification of judgments affects the performance of evaluation metrics. We note that these modifications in relevance judgements are only applied for measuring $\textit{FD}$ and nDCG@10 is measured with all the judged documents without any modification. 

We plotted the nDCG@10 on the x-axis and the $\textit{FD}$ with balanced and sparsified judgments on the y-axis of each sub-figure in Figure \ref{fig:trec}, for all the runs submitted to TREC DL19 (first row) and TREC DL20 (second row). Consistent with our previous experiments, we observe a highly linear relationship between the two metrics. We also provide the Kendall $\tau$ correlation under each sub-figure. For instance, when sparsifying the labels and considering only one relevant judged item per query, we obtain a Kendall $\tau$ correlation of -0.836 for TREC DL2019 and -0.867 for TREC DL2020, between nDCG@10 and $\textit{FD}@10$ of each dataset.

In addition, we present the Kendall Tau correlation between nDCG when using full relevance judgments versus randomly selecting a maximum of $N$ relevant judgments, where $N$ could be 1, 5, or 10, as illustrated in Figure \ref{fig:trec}. It is worth noting that while FD (as demonstrated in Figure \ref{fig:distribution}) exhibits a higher degree of robustness when evaluated with sparse labels, nDCG is not as resilient concerning the chosen relevant judged document (qrel). This is because FD computes its metrics over the distribution of all queries, contributing to a more stable evaluation performance. On the contrary, NDCG with sparse labels tends to be considerably noisy and heavily dependent on which document is selected as the ``one relevant document'' per query, leading to significant variations in the results.
In the Table \ref{tab:new_addedd} , we present the Kendall Tau correlation between nDCG with full relevance judgements and nDCG when choosing 1, 5, or 10 random relevant documents. These results highlight the sensitivity of nDCG to the choice of relevant documents, especially when only a limited number of relevant documents are considered.

\begin{table}
    \centering
    \begin{tabular}{cccc}
     \hline \hline 
      Dataset   & 10 qrels &  5 qrels & 1 qrel \\ \hline 
     Trec-DL-2019 &   0.796 &  0.784 & 0.594 \\
    Trec-DL-2020 &   0.918 &  0.891 &  0.863 \\  \hline \hline
    \end{tabular}
    \caption{Kendall Tau correlation between nDCG measured with full relevance judgements and sparsified relevance judgements. }
    \label{tab:new_addedd}
\end{table}
The experiments on the TREC DL datasets highlight two key points. First, unlike using the Fréchet Inception Distance to evaluate the quality of generated images in text-to-image generation tasks, where a large number of data points (in the order of thousands) are required for the evaluation to be valid, we demonstrated that even with a smaller number of queries (around 40-50), $\textit{FD}$ is capable of distinguishing the performance of different rankers \cite{Kynkaanniemi2022,heusel2017gans}. Second, $\textit{FD}$ is not sensitive to the sparsity of the ground truth labels and it performs well with both sparse and more complete labels. It is not affected by the number of judgments, as evidenced by the fact that the performance did not differ greatly when increasing the number of relevant judged items. However, for TREC DL2019, we observed a small drop in correlation by increasing the number of relevant judgments. Further exploration revealed that a higher number of relevant judgments in TREC 2019 resulted in a higher usage of level 2 relevance judgments (highlight relevant) instead of level 3 judgments (perfectly relevant). Consequently, we suggest that $\textit{FD}$ may be more sensitive to the quality of relevant judged items rather than the quantity.
Overall, in response to \textbf{RQ2}, we find that $\textit{FD}$ works well when using comprehensive labels, and consistent with the findings in Section \ref{results1}, sparsifying the labels does not compromise the quality of assessment.

\section{Assessing Unlabeled Retrieved Results} 
\label{URI}
Here, we undertake an evaluation of different IR systems under an extremely challenging case of assessing unlabeled retrieved results.  This scenario presents a situation where each query is assumed to have mostly only one relevant item, and the \textit{relevant judged items are not included in the top-$k$ results}. 
Our objective is to investigate the effectiveness of the Fréchet Distance in assessing the top-$k$ Unlabeled Retrieved Results (URR) when no judgments are available for any of the top-$k$ retrieved items.  
This is particularly valuable considering the high cost and limited availability of labeled data, which often exhibit sparsity. Previous research has demonstrated that as rankers improve in performance, they tend to retrieve previously unseen content that may be highly relevant to the original query \cite{arabzadeh2022shallow}. If Fréchet Distance is capable of evaluating the retrieved results in such cases, it would be a valuable tool for assessing the relevance of unlabeled data and even beyond that, for evaluating generative-based responses. 

We measure the $\textit{FD}$ between one set consisting of the relevant judged items per query and the other set consisting of the top-$k$ \textit{unjudged} retrieved item for each query. In other words, we scan down the ranked list and retain the first $k$ unjudged document to assess. This is an interesting aspect to study because traditional IR metrics such as MRR, nDCG, and MAP rely on the presence of relevant items in the retrieved list and would assign a performance score of zero in cases where no relevant items are retrieved. They do not account for unjudged documents. We argue that by utilizing the $\textit{FD}$ metric, we can capture the similarity between unjudged retrieved items and the limited set of judged examples and measure the performance of the retriever based on this value. 

\begin{table}[]
\caption{Performance of different retrievers in terms of MRR@10 as well as Fréchet distance $\textit{FD}$ assuming under Unlabeled Retrieved Results (URR) setting. We note that the MRR@10 is measured on the original ranked list since with URR setting, all the retrievers would obtain MRR@10 equals to zero.  A smallest Fréchet distance corresponds to better performance.}
\scalebox{0.78}{
\begin{tabular}{llrrr}
\hline\hline
& & &  \multicolumn{2}{c} {URR}  \\ \cline{4-5}
Category & Method & \multicolumn{1}{l}{MRR@10} & \multicolumn{1}{l}{$\textit{FD}@1$} & \multicolumn{1}{l}{$\textit{FD}@10$} \\ \hline
\multirow{3}{*}{Sparse} & BM25 & 0.187 & 8.634 & 4.705 \\ 
 & DeepCT & 0.242 & 4.183 & 2.591 \\ 
 & DocT5 & 0.276 & 4.066 & 2.290 \\ \hline
\multirow{5}{*}{Dense} & RepBERT & 0.297 & 2.701 & 1.364 \\ 
 & ANCE & 0.330 & 2.353 & 1.126 \\ 
 & SBERT & 0.333 & 2.266 & 1.156 \\ 
 & ColBERT & 0.335 & 2.308 & 1.115 \\ 
 & ColBERT V2 & 0.344 & 2.352 & 1.121 \\ \hline
\multirow{2}{*}{\begin{tabular}[c]{@{}l@{}}Trained\\  Sparse\end{tabular}} & UniCOIL & 0.351 & 2.302 & 1.128 \\ 
 & SPLADE & 0.368 & 2.300 & 1.117 \\ \hline
\multirow{2}{*}{\begin{tabular}[c]{@{}l@{}}Hybrid\\ (BM25)\end{tabular}} & ColBERT-H & 0.353 & 2.399 & 1.115 \\ 
 & ColBERT V2 -H & 0.368 & 2.365 & 1.142 \\ \hline\hline
\end{tabular}}
\label{tab:urr}

\end{table}

The results of this experiment are reported in Table \ref{tab:urr} with two cut-offs of ``$\textit{FD}@10$'' and ``$\textit{FD}@1$''. Even when no judged documents appear in the top-$k$, $\textit{FD}$ is still able to quantify the performance of the retriever. This capability is not present in traditional metrics. 
For instance, when there are no relevant judged items retrieved in the ranked list, $\textit{FD}@1$ quantifies the performance of BM25 as 8.634, whereas the performance for ColBERT is measured as 2.308. This indicates that even without relevant judged items, $\textit{FD}$ is capable of determining that ColBERT performs better than BM25.

This experiment demonstrates that, unlike traditional IR metrics, $\textit{FD}$ is not sensitive to the labeled documents themselves. Indeed, the Fréchet Distance is not reliant on the exact positioning of the relevant judged document in the ranking. Instead, it focuses on measuring the similarity between the retrieved items and the relevant judged documents. This characteristic makes it particularly valuable for evaluating scenarios with extremely sparse labels, even in cases where the rankers do not retrieve the labeled data. In response to \textbf{RQ3}, the Fréchet Distance enables assessment of the remaining unlabeled data, offering valuable insights into their relevance. \textit{In contrast, traditional IR metrics would be unable to provide any insights without retrieving the labeled documents.}

\begin{table}[]
\caption{Kendall $\tau$ correlation between different evaluation metrics over the 12 retrieval methods. URR stands for ``Unlabeled Retrieved Results'' and refers to experimental results from section \ref{URI}. All the correlations are statistically significant with p-value $<$ 0.05}
\scalebox{0.69}{
\begin{tabular}{p{2.3cm}rrrrr}
\hline\hline
 & MRR@10 & $\textit{FD}@1$ & \begin{tabular}[c]{@{}c@{}}$\textit{FD}@1$ \\ URR\end{tabular} & $\textit{FD}@10$ & \begin{tabular}[c]{@{}c@{}}$\textit{FD}@10$ \\ URR\end{tabular} \\ \hline
MRR@10 & 1 & -0.473 & -0.545 & -0.788 & -0.636 \\ 
$\textit{FD}@1$ & -0.473 & 1 & 0.687 & 0.443 & 0.290 \\ 
$\textit{FD}@1$-URR & -0.545 & 0.687 & 1 & 0.636 & 0.485 \\
$\textit{FD}@10$ & -0.788 & 0.443 & 0.636 & 1 & 0.848 \\ 
$\textit{FD}@10$-URR & -0.636 & 0.29 & 0.485 & 0.848 & 1 \\ \hline\hline
\end{tabular}
}
\label{kendal}
\end{table}

\section{Further analysis}
\subsection{Correlation with IR Evaluation Metrics}
\label{corr}
We aim to examine the correlation between the $\textit{FD}$ measure and 
traditional IR evaluation metrics. To achieve this, we calculate the ranked-based Kendall $\tau$ correlation, for each pair of metrics in Table \ref{mainresults} and Table \ref{tab:urr} on the performance of the 12 retrievers introduced earlier and report the results in Table \ref{kendal}. This set of evaluation metrics includes MRR@10, $\textit{FD}$ at cut-offs 1 and 10 (Section \ref{results1}) and $\textit{FD}$ at cut-offs 1 and 10 under URR setting when no labeled data is retrieved (Section \ref{URI}).
 As anticipated and illustrated in Figure \ref{fig:trec}, $\textit{FD}$ exhibits a negative correlation with MRR, as a lower $\textit{FD}$ value indicates better performance. Among these correlations, $\textit{FD}@10$ shows the highest absolute correlation with MRR@10 i.e., a correlation of -0.788. We suggest that this is because $\textit{FD}$ operates based on the distribution of embedded representations of documents, which has shown to work most stably when the number of samples increases \cite{DBLP:journals/corr/abs-1911-07023,binkowski2018demystifying}. 
More interestingly, $\textit{FD}@1$  and $\textit{FD}@1$ with Unlabeled Retrieved Results (URR),  obtain a correlation coefficient of 0.687. Similarly, the correlation between $\textit{FD}@10$ (Fréchet Distance at 10) and $\textit{FD}@10$ with unlabeled retrieved items was found to be 0.848. The high correlation between evaluating the original retrieved results vs without having any judged retrieved results further validates the findings presented in sections  \ref{results1} and \ref{URI}.The Fréchet Distance not only exhibits a high correlation with traditional IR metrics but also demonstrates its capability in assessing unlabeled retrieved items.
These experiments let us answer \textbf{RQ4} that $\textit{FD}$ shows a notable correlation with traditional IR metrics. These properties increase the reliability of using $\textit{FD}$ for assessing IR systems.

\begin{table}[]
\caption{Comparison of the performance of different retrievers when assessing with MRR@10 and $\textit{FD}@10$ on MS MARCO dev set 
With DistilBERT fine-tuned on MSMARCO as well as DistilBERT without any fine-tuning. DistilBERT fine-tuned on MSMARCO shows  $-0.788$ Kendall $\tau$ correlation with MRR@10 and DistilBERT without any fine-tuning shows $-0.739$ Kendall $\tau$ correlation with MRR@10.  }
\scalebox{0.63}{
\begin{tabular}{llrrr}
\hline\hline
 &  & \multicolumn{1}{l}{} & \multicolumn{2}{c}{$\textit{FD}@10$} \\ \cline{4-5}
\multicolumn{1}{l}{Category} & \multicolumn{1}{l}{Method} & \multicolumn{1}{l}{MRR@10} & \multicolumn{1}{l}{\begin{tabular}[c]{@{}c@{}}DistilBERT\\ MSMARCO\end{tabular}} & \multicolumn{1}{c}{\begin{tabular}[c]{@{}c@{}}DistilBERT\\ No Fine-tuning\end{tabular}} \\ \hline
\multicolumn{1}{l}{} & \multicolumn{1}{l}{BM25} & 0.187 & \multicolumn{1}{r}{0.590} & 4.410 \\
\multicolumn{1}{l}{} & \multicolumn{1}{l}{DeepCT} & 0.242 & \multicolumn{1}{r}{0.412} & 2.354 \\
\multicolumn{1}{l}{\multirow{-3}{*}{Sparse}} & \multicolumn{1}{l}{DocT5} & 0.276 & \multicolumn{1}{r}{0.331} & 2.050 \\ \hline
\multicolumn{1}{l}{} & \multicolumn{1}{l}{RepBERT} & 0.297 & \multicolumn{1}{r}{0.159} & 1.223 \\
\multicolumn{1}{l}{} & \multicolumn{1}{l}{ANCE} & 0.330 & \multicolumn{1}{r}{0.121} & 0.995 \\
\multicolumn{1}{l}{} & \multicolumn{1}{l}{SBERT} & 0.333 & \multicolumn{1}{r}{0.132} & 1.008 \\
\multicolumn{1}{l}{} & \multicolumn{1}{l}{ColBERT} & 0.335 & \multicolumn{1}{r}{0.117} & 0.980 \\
\multicolumn{1}{l}{\multirow{-5}{*}{Dense}} & \multicolumn{1}{l}{ColBERT V2} & 0.344 & \multicolumn{1}{r}{0.118} & 0.982 \\ \hline
\multicolumn{1}{l}{} & \multicolumn{1}{l}{UniCOIL} & 0.351 & \multicolumn{1}{r}{0.123} & 0.980 \\
\multicolumn{1}{l}{\multirow{-2}{*}{\begin{tabular}[c]{@{}l@{}}Trained\\  Sparse\end{tabular}}} & \multicolumn{1}{l}{SPLADE} & 0.368 & \multicolumn{1}{r}{0.120} & 0.964 \\ \hline
\multicolumn{1}{l}{} & \multicolumn{1}{l}{ColBERT-H} & 0.353 & \multicolumn{1}{r}{0.116} & 0.973 \\ 
\multicolumn{1}{l}{\multirow{-2}{*}{\begin{tabular}[c]{@{}l@{}}Hybrid\\ (BM25)\end{tabular}}} & \multicolumn{1}{l}{ColBERT V2 -H} & 0.368 & \multicolumn{1}{r}{0.126} & 0.998 \\ \hline\hline
\end{tabular}}
\label{BERT-raw}
\end{table}

\subsection{Impact of Document Representation}
\label{robustnessembd}
Here, we examine the robustness of the Fréchet Distance metric for assessing IR systems with respect to the underlying language model to embed the retrieved documents and relevance judgments. We aim to investigate how the choice of language model impacts the quality of evaluating IR systems using the Fréchet Distance measure considering this change would vary the document feature vectors.
For previous experiments, we utilized a language model that was fine-tuned on the MS MARCO dataset for ranking tasks. However, now we study how the results would be impacted if we were to embed the retrieved documents and ground truth in a different space. 
As such, we present the same results as in Table \ref{mainresults}, using DistilBERT embeddings fine-tuned on the MSMARCO training set as well as the same results with a DistilBERT without any fine-tuning. This analysis aims to investigate whether a general-purpose language model can capture the necessary information for accurate assessment, or if a language model specifically fine-tuned for ranking tasks in retrieval is required. Table \ref{BERT-raw} displays the obtained results.
Surprisingly, we observe that changing the language model from a fine-tuned ranking model to a raw, unfine-tuned BERT model does not substantially impact the assessment outcomes. The $\textit{FD}$ metric remains capable of effectively evaluating the performance of various retrieval methods. For example, from Table \ref{BERT-raw}, and under ``DistilBERT No fine-tuning'' column, we observe that BM25 achieves an $\textit{FD}@10$ score of $4.410$, whereas COLBERT, which is expected to be a better model, achieves a score of $0.980$.

The correlation between $\textit{FD}@10$ and MRR@10 when using DistilBERT without any fine-tuning, is -0.739. Comparatively, when using fine-tuned DistilBERT (as shown in Table \ref{kendal}), the correlation IS -0.788. As such, having a fine-tuned language model specifically for ranking task can improve the correlation with traditional IR metrics. However,  even without fine-tuning, $\textit{FD}$ still demonstrates promising performance.
Overall, the results indicate that $\textit{FD}$ remains effective in evaluating the quality of retrieved results, even when employing a general-purpose language model without fine-tuning.
Lastly, with respect to \textbf{RQ5}, we note that $\textit{FD}$ shows promising robustness w.r.t  the document embedding representation.

\section{Conclusion and Future work}

In this paper, we leverage Fréchet Distance to address the challenges of evaluating IR systems with sparse labels. We measure the similarities between the embedded representation of retrieved results as well as the limited available relevant judged documents using Fréchet Distance. 
Through experiments conducted on datasets with sparse and more complete ground truth labels, including the MS MARCO DEV dataset and the TREC Deep Learning Track datasets
, we demonstrated  the effectiveness of the Fréchet Distance in evaluating IR systems. our findings suggest that the Fréchet Distance has significant implications for evaluating IR systems in real-world settings where obtaining comprehensive ground truth labels can be challenging and expensive. We believe that future research could utilize the Fréchet Distance to evaluate different generative models, expanding the scope of evaluation in IR systems. As such, it allows for having the generated results compared with the retrieved results in the same playground.

\section{Limitations}
While our study provides valuable insights into the effectiveness of the Fréchet Distance in evaluating IR systems with sparse labels, there are a few limitations that should be acknowledged. First, unlike traditional IR evaluation metrics, the Fréchet Distance is not applicable to individual queries and can only be used with sets of queries. Further exploration is needed to understand how the sample size of the queries affects the quality of the assessment. 
Second, the Fréchet Distance assumes that the two distributions follow a multivariate normal distribution. Lastly, it is important to note that the Fréchet Distance is an unbounded metric, and its range varies depending on the dataset's characteristics and the number of samples under investigation.
Building upon the findings of this study,


\bibliography{acl}
\bibliographystyle{acl_natbib}

\end{document}